\documentclass[a4paper,11pt]{article}
\usepackage{jheppub} 
\usepackage{lineno}
\usepackage{color}

\definecolor{darkgreen}{RGB}{50,150,0}

\definecolor{darkred}{RGB}{200,20,0}

\title{Astrophysical Constraints on Decaying Dark Gravitons}

\author[a]{Jamie A.P. Law-Smith,}
\author[b]{Georges Obied,}
\author[c]{Anirudh Prabhu,}
\author[d]{Cumrun Vafa}
\affiliation[a]{Department of Astronomy and Astrophysics, University of Chicago, Chicago, IL 60637, USA}
\affiliation[b]{Rudolf Peierls Centre for Theoretical Physics, University of Oxford, Oxford OX1 3PU, United Kingdom}
\affiliation[c]{Princeton Center for Theoretical Science, Princeton University, Princeton, NJ 08544, USA}
\affiliation[d]{Department of Physics, Harvard University, Cambridge, MA 02138, USA}

\abstract{In the dark dimension scenario, which predicts an extra dimension of micron scale, dark gravitons (KK modes) are a natural dark matter candidate. In this paper, we study observable features of this model. In particular, their decay to standard matter fields can distort the CMB and impact other astrophysical signals.  Using this we place bounds on the parameters of this model.  In particular we find that the natural range of parameters in this scenario is consistent with these constraints and leads to the prediction that the mean mass of the dark matter today is close to a few hundred keV and the effective size of the extra dimension is around $1 - 30 \;\mu\mathrm{m}$. }

\begin{document}

\maketitle
\flushbottom

\color{black}

\section{Introduction}

The dark dimension scenario~\cite{Montero:2022prj}, which is motivated by the smallness of the dark energy density in the context of the distance conjecture of the Swampland program (see e.g.~\cite{Agmon:2022thq}), combined with observational data, has led to the prediction of one mesoscopic dimension whose length scale is in the range $0.1 \ \mu\rm{m} < \ell <30 \ \mu\rm{m}$.  The mesoscopic extra dimenion in turn leads to the prediction of the existence of a light tower of graviton Kaluza-Klein (KK) modes, the ``dark gravitons'' (DGs), which provide an unavoidable contribution to dark matter. In previous work~\cite{Gonzalo:2022jac}, it was shown that the DGs can in fact constitute all of the dark matter density we observe today.  
This scenario has thus led to a unification of the dark sector where the smallness of the dark energy leads to an unavoidable candidate for the dark matter.

Dark matter is created from the emission of DGs from the standard model (SM) fields in thermal equilibrium beginning at a temperature of $\sim$ GeV. The KK modes thus produced have a mean mass of $\sim$ GeV and start decaying to lower KK modes as time evolves, decreasing their mean mass with time.  Unlike the usual scenarios for dark matter which involve a single or a few stable particles\footnote{See for example~\cite{Kolb:2023dzp} for a discussion of a single massive spin-2 particle and its gravitational production in cosmology.}, this scenario involves a tower of slowly decaying particles, leading to a changing composition over time.
This type of model of dark matter naturally arises more broadly in the context of string theory and has been called the `dynamical dark matter'~\cite{Dienes:2011ja,Dienes:2011sa}.

The cosmological aspects of the dark dimension scenario, which were considered in~\cite{Gonzalo:2022jac} involve the production of DGs, which is unavoidable due to the universality of gravitational interactions. This scenario passes a number of non-trivial checks. In addition to inter-tower decays of dark gravitons, they can also decay back to SM fields such as photons and fermions, which can leave a footprint in the cosmic microwave background (CMB) and contribute to the extragalactic background radiation observed over a wide range of frequencies. It is the purpose of this paper to constrain some parameters in this scenario using observational data on the CMB as well as galactic and extragalactic $X$-ray emission. We carefully study these imprints and use the observational data to place constraints on how strongly DGs can couple to the SM brane. Under the assumption that dimensionless coupling constants of decay are natural, we find that the dark matter mass today should be below a few hundred keV, which is consistent with the rough estimates anticipated in~\cite{Gonzalo:2022jac}.

In the rest of the paper we discuss these constraints starting with early universe constraints from the CMB, extragalactic background and Lyman-$\alpha$. Amongst these, the CMB places the strongest bound. We then discuss the ones based on late universe physics including the Milky Way, and 511 keV line.  In the last section of the paper we summarize our conclusions for the Dark Dimension scenario and discuss the natural parameter range showing that it lies just beyond what current data can probe.

\section{Dark Dimension Phenomenology}
\label{sec:DDPheno}

\begin{figure}
    \centering    \includegraphics[width=0.75\linewidth]{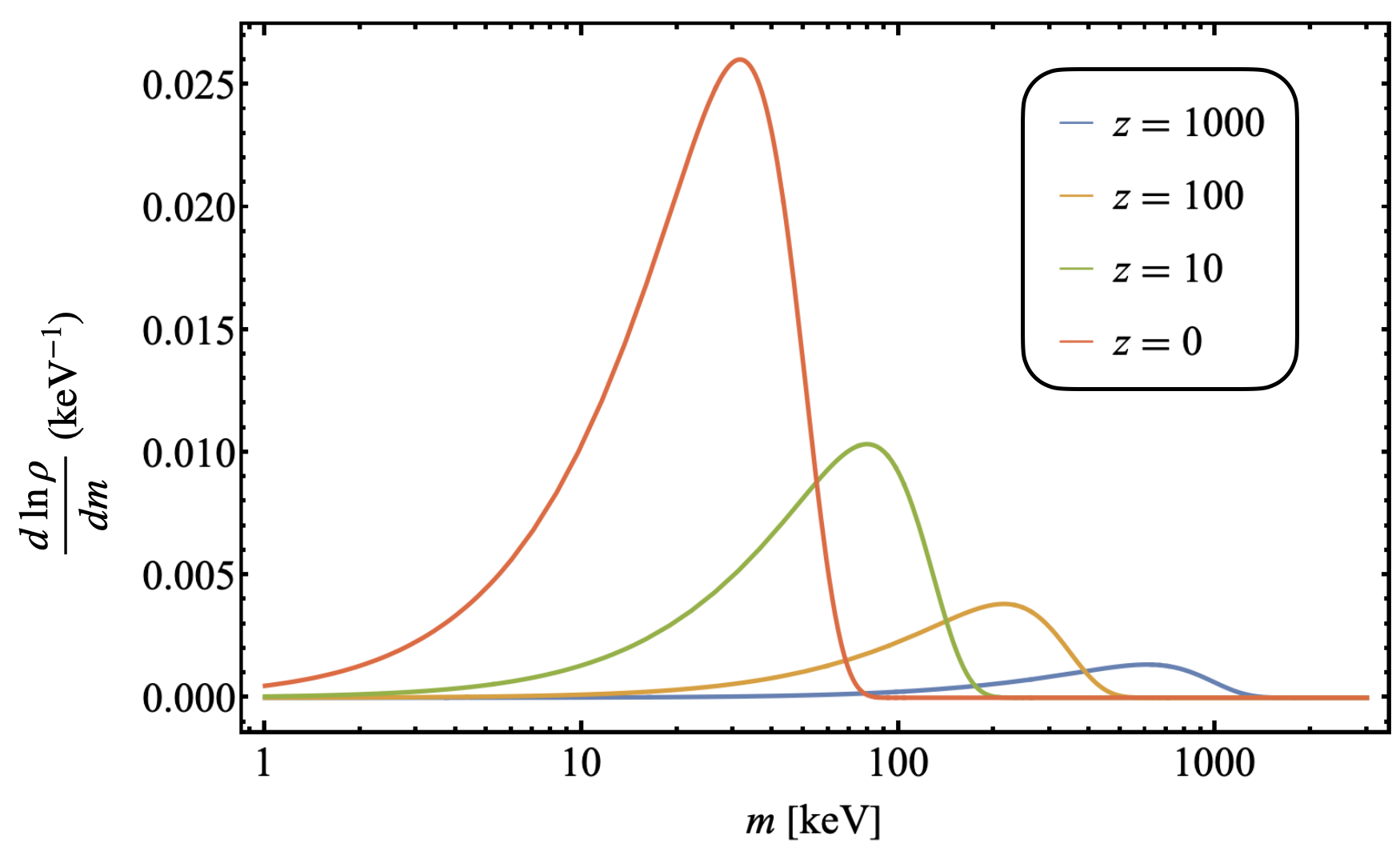}
    \caption{A representative distribution of dark gravitons among KK states as a function of redshift.}
    \label{fig:MassSpectrum}
\end{figure}

In the dark dimension scenario, KK modes of the graviton in one mesoscopic extra dimension, play the role of dark matter. The phenomenology of the dark dimension has been studied in various recent works~\cite{Anchordoqui:2022svl,Anchordoqui:2023oqm,Anchordoqui:2023tln,Cribiori:2023swd,Blumenhagen:2022zzw}. In this work we focus on the scenario described in~\cite{Gonzalo:2022jac} where dark matter consists of a tower of excitations of the graviton in the 5th dimension. The dark matter is produced by the coupling of the matter fields in the visible sector with the 5d graviton.  The cosmological scenario starts with the matter fields in equilibrium with initial temperature in the GeV range. These are localized excitations in the 5th dimension which lead to the creation of dark gravitons initially in the GeV mass range. Due to weak gravitational couplings, the tower decays gradually to lower the mass of the dark matter, leading to a lighter dark matter profile. The decay is sufficiently slow that by the time we get to the $T_\gamma\sim 1\;\mathrm{eV}$ the dark tower has mass in the MeV range begins to dominate the energy density in the universe, before the dark energy takes over.  The rate of decrease of the mass is determined by intra-tower decay rate given by:
\begin{align}
        \Gamma^{\text{tot}}   = \beta^2 \frac{m^{7/2}\delta^{3/2}}{m_{KK}^{1/2}M_P^2}  
        \label{eq:totWidth},
\end{align}
where $m \sim n m_{KK}$ is the mass of the state with quantum number $n$, $\delta$ is a measure of the smoothness of the extra dimension and determines the range of violation of the KK quantum number $n$ in the decays. $\beta$ is an $O(1)$ parameter that controls the strength of the intra-tower decay amplitudes which correlates with the amplitudes on inhomogeneities in the 5th dimension. For a more extensive discussion of these parameters, see~\cite{Gonzalo:2022jac} or section~\ref{sec:results} below.
This decay rate was estimated in~\cite{Gonzalo:2022jac} assuming that the decay to all lower modes proceeds gravitationally.  Using this decay rate, one sees that the central mass of the dark sector scales with time as (see Figure~\ref{fig:MassSpectrum})
\begin{align}
m_{DM}(t)\sim m(t_0)\bigg
(\frac{t}{t_0}\bigg)^{-2/7}.
\label{eq:massDM}
\end{align}

In particular, this decay mechanism ensures that the lifetime of dark matter at any moment in cosmic history is comparable to the Hubble time at that moment. This is easy to see from equations \eqref{eq:totWidth} and \eqref{eq:massDM} which immediately imply that the lifetime $\tau \propto \Gamma^{-1} \propto t \propto H^{-1}$. In fact, this is how the time-dependence~\eqref{eq:massDM} is deduced. This dependence on the Hubble rate of the decay width between dark sector fluids has been studied extensively in the literature albeit from a more phenomenological perspective (see~\cite{Valiviita:2008iv} and references therein). Our model provides a physical mechanism to realize such time-dependent decay rates. Without such a mechanism, it would be difficult to explain~\cite{Valiviita:2008iv} why a local quantity (the decay rate $\Gamma$) depends on the cosmological expansion rate $H$. One potential way to link the decay and expansion rates was proposed in~\cite{Bjaelde:2012wi}, where the DM candidate $\phi$ is a coherent scalar field oscillating around a minimum and decaying to a nearly-massless fermion $\psi$ with Yukawa coupling $\phi \overline{\psi}\psi$. The fermion is produced by a parametric resonance effect when its effective mass crosses zero. The expansion of the universe affects the background solution of the dark matter field $\phi$, leading to a time-dependence in the $\phi$-to-$\psi$ decay rate. To our knowledge, however, no such realization of $\Gamma \propto H$ exists for particle dark matter candidates (as opposed to coherent fields). The time-dependence of the decay rate may be linked to new and interesting phenomenology (see for instance~\cite{Pandey:2019plg} for an application of decaying dark matter with $\Gamma \propto H$ to the $H_0$ and $\sigma_8$ tensions).

The DGs also decay back to the SM fields.  In particular their decay rate to photons is given by~\cite{Han:1998sg,Hall:1999mk}

\begin{equation} \Gamma_{{\gamma \gamma}}=\frac{{\lambda}^2m^3}{{80\pi M_P^2}},
    \label{decaytophoton}
    \end{equation}
and to $e^-e^+$ the decay rate is

\begin{equation} \Gamma_{{e^+e^-}}=\frac{{\lambda}^2m^3}{{160\pi M_P^2}} \left(1 - 4 \frac{m_e^2}{m^2}\right)^{3/2} \left(1 + \frac83 \frac{m_e^2}{m^2}\right).
    \label{decaytoelectron}
    \end{equation}
where $\lambda\sim O(1)$ measures the value of the dark graviton wave function at the SM brane and is expected to be order 1. The normalization of the decay rates is written so that $\lambda$ is exactly one when the wave function at SM brane is the usual $1/\sqrt{L_5}$ where $L_5$ is the length of the 5-th dimension.    
The main aim of this paper is to find a bound on the parameters of the model using the decay back to SM fields and the imprint this leaves on CMB and $X$-ray backgrounds. In particular we will produce an exclusion plot in the $\lambda-m$
plane.  If we take $\lambda \sim O(1)$
this can be interpreted as a bound on the mass of dark matter today. In addition, using input from other observables (see section~\ref{sec:results}), this bound in turn leads to constraints on the parameters $\delta,\beta, m_{KK}$ of the model.
Note that our model depends on the four parameters: $\lambda,\beta,\delta,m_{KK}={1/L_5}$,
the first three of which are expected to be $\mathcal{O}(1)$ numbers and $m_{KK}>6.6\ \mathrm{meV}$ \cite{Tan:2016vwu,Lee:2020zjt}. This latter number is an experimental bound from the lack of deviation from Newton's inverse square law up at distances as short as $30\ \mu m$.

The dark dimension scenario gives a concrete realization of our dark matter sector and its evolution throughout cosmic history. One feature of the model, the fact that DGs can decay to SM particles, means that current and near future experiments can detect these graviton decay products to confirm or rule out the model. It turns out\footnote{The other potentially relevant channel is decay to neutrinos. This has very little impact on cosmology in our model. First, the decay to neutrinos in the early universe can change the energy density in radiation, parameterized by $N_\mathrm{eff}$. Any additional energy density in radiation is bounded above by the energy loss from DM which can be seen to be small at all epochs~\cite{Gonzalo:2022jac}. The other potentially relevant effect is if the decays can produce an asymmetry in the neutrino vs. anti-neutrino abundance which could ultimately lead to changing the Helium abundance~\cite{Serpico:2005bc} for instance. However, decays in our model cannot affect the neutrino asymmetry and therefore related effects are absent. } that only decays to photons and electrons will have (small) effects on astrophysics and cosmology. We discuss these effects in turn before presenting a more detailed analysis in the following subsections.

We start with the CMB measurements from {\it Planck} which provide a precision probe of the early universe and can in principle be sensitive to new physics that alters the photon-baryon fluid between today and the surface of last scattering. In particular, the damping of the CMB power spectrum is affected by the integrated optical depth along the line of sight (see for example~\cite{dodelson2020modern}). Additional energy injections (from decaying DM for example) in the redshift range $20 \lesssim z \lesssim 1100$ can ionize Hydrogen and Helium and increase the optical depth to recombination causing a stronger damping of the CMB power spectrum on small scales. The absence of such a suppression in the observed spectrum places an upper bound on the energy injection from dark matter decays. For our model, where the decay width is given in~\eqref{decaytophoton}--\eqref{decaytoelectron}, this will be an upper bound on $\lambda$ for each value of the DG mean mass $m_\mathrm{today}$. Given the dependence on $\lambda$ and $m$ in the decay width, we have:
\begin{align}
    \label{eq:CMBBound}
    \lambda \lesssim 0.1 \times \left(\frac{m_\mathrm{today}}{100 \;\mathrm{keV}}\right)^{-3/2} \qquad \text{(from the CMB)}.
\end{align}
where the numerical prefactor is determined from the data analyses discussed below. 

Similarly, energy injection can increase the intergalactic medium (IGM) temperature after the epoch of reionization (redshifts $2 \lesssim z\lesssim 6$). The latter temperature and its evolution at low redshift can be measured from the Lyman-$\alpha$ flux power spectrum~\cite{Schaye:1999vr,Becker:2010cu}. A decaying DM component can heat the IGM and alter this temperature profile. The data then constrains any such heating sources and for our model implies
\begin{align}
    \lambda \lesssim 0.1\left(\frac{m_\mathrm{today}}{1\;\mathrm{MeV}}\right)^{-3/2} \qquad \text{(from Ly-$\alpha$)}.
\end{align}

While the CMB provides the strongest constraint on our model, we also check various other direct signals of DG decay which also imply constraints on $\lambda$ as a function of the DG mean mass today $m_\mathrm{today}$. For instance, the decay of DM to photons throughout cosmic history contributes to the diffuse extragalactic background light observed by various satellites such as \cite{Cappelluti:2017miu,gendreau1995asca,Gruber:1999yr,kinzer1997diffuse,churazov2007integral,Revnivtsev:2003wm,Ajello:2008xb,fukada1975energy,watanabe1999diffuse,weidenspointner2000cosmic}. There is also a galactic $X$-ray signal from similar decays to photons. These constraints are weak for the mass range of interest and allow $\lambda > 1$ for $m_\mathrm{today}$ less than a few MeV. However, the decay of DGs to electron-positron pairs that contribute to the 511 keV  $\gamma$-ray signal from our galaxy provides a stronger constraint. The requirement that the contribution from this decay does not exceed the observed value implies
\begin{align}
    \lambda \lesssim 0.01 \left(\frac{m_\mathrm{today}}{1 \;\mathrm{MeV}}\right)^{-3/2}
    \qquad \text{(for $m_\mathrm{today} \gtrsim 500\;\mathrm{keV}$, from 511 keV line)}.
\end{align}
Recalling that the coupling $\lambda$ is an averaged phenomenological parameter that captures the overlap of the bulk graviton wave function with the brane and is expected to be $\mathcal{O}(1)$ the above two constraints imply that a natural mass range for the DM is less than about $100 \;\mathrm{keV}$ today.

\begin{figure}
    \centering
    \includegraphics[width=\linewidth]{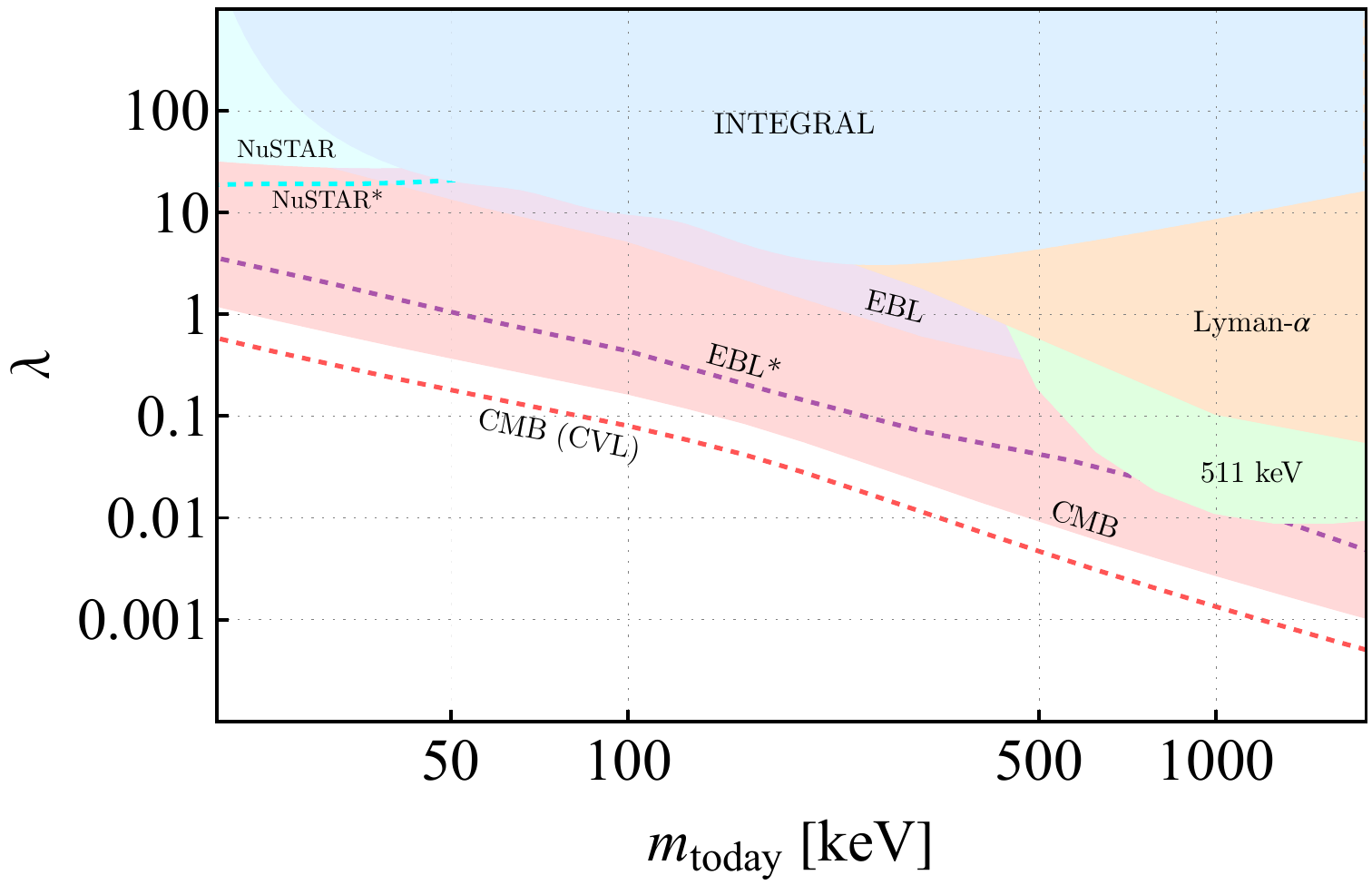}
    \caption{Constraints on the strength of coupling of DGs to the SM brane from CMB anisotropies measured by \emph{Planck} (light red, shaded) and by a future cosmic-variance-limited (CVL) measurement (red, dashed). We have also included constraints from extra-galactic background light (EBL) with a conservative bound shown in the shaded purple region, and an optimistic bound based on understanding of astrophysical backgrounds.  For this latter bound we assume astrophysical effects can explain all the EBL signal and ensure that DG decay produces an effect smaller than the error bars.  This is represented by the purple dashed line (EBL$^*$). Also shown are constraints from $X$-ray measurements by the INTEGRAL satellite (light blue, shaded), NuSTAR galactic center observations (conservative constraints in shaded cyan and optimistic constraints derived from subtracting known astrophysical sources as the dashed line and labeled with an asterisk), Lyman-$\alpha$ measurements of the IGM temperature (light orange, shaded), and the 511 keV line (light green, shaded). Constraints are quoted in terms of the peak of the mass distribution of the DG states at the present day.}
    \label{fig:Constraints}
\end{figure} 

\section{Constraints}

\subsection{Early Universe Constraints}
\subsubsection{CMB Constraints}
\label{sec:CMB}

The CMB gives a snapshot of the Universe at the time ($z_{\rm rec} \sim 1100$) when the free electron fraction dropped precipitously, allowing for photons to free-stream. The spectrum of these photons is very well-approximated by a black-body. Deviations in temperature of this black body as a function of observation angle in the sky (temperature anisotropies) have been measured with exquisite precision by a number of experiments including WMAP\cite{WMAP:2012fli}, SPT \cite{SPT-3G:2022hvq}, ACT \cite{ACT:2020frw}, and the \emph{Planck} satellite~\cite{Planck:2018nkj}. Dark matter decaying between recombination and reionization, when radiation from the first stars and galaxies reionized the the Universe, injects high-energy particles into the plasma that can ionize neutral Hydrogen and Helium and heat or excite the gas, leading to modifications to the CMB anisotropy spectrum. In this section we use measurements of the CMB anisotropy spectrum to constrain energy injection from decaying DGs. 

As DGs decay, they inject high energy particles into the SM plasma, with a volumetric rate given by 

\begin{align} \label{eqn:injectionrate}
    \left( {\frac{d E}{dV dt}} \right)_{{\rm inj},i} = \displaystyle\int  e^{- \Gamma_{{\rm KK}, i}(m) t} {\Gamma_{{\rm KK}, i}(m) } \rho_{{\rm DM}, 0} \left( {\frac{d \ln \rho}{dm}}\right) (1+z)^3 dm,
\end{align}

\noindent where the index $i$ refers to the SM states to which the DGs decay and $\rho_{{\rm DM}, 0}$ is the present-day dark matter density. As discussed in the previous section, we assume the KK spectrum to be continuous given that its mass splitting is much smaller than the typical masses that are of interest in this paper. In~\eqref{eqn:injectionrate}, we sum the energy injection over all KK states. This injected energy is partitioned into several channels, $c \in$ \{H ionization, He ionization, heating, excitation, continuum photons\}. At a given redshift, the amount of energy deposited into each of these channels depends on the particle injected (e.g. $e^\pm$ and photons), the redshift of injection, and the conditions of the universe at that redshift. It is particularly important to note that energy deposition is not necessarily prompt; there may be a significant delay between injection and deposition. The energy deposition history can be related to the injection history through the ``$f$-curves'', defined as

\begin{align} \label{eqn:depositionrate}
    \left( {\frac{d E}{dV dt}} \right)_{{\rm dep},c} = f_c(z) \left( {\frac{d E}{dV dt}} \right)_{{\rm inj}},
\end{align}

\noindent where $(dE/dVdt)_{{\rm dep}, c}$ is the energy deposition rate into channel $c$. Given an arbitrary injection history, $f_c(z)$, can be computed using the methods described in \cite{Slatyer:2015kla}. The functions $f_c(z)$ fully characterize the effect of energy injection from DG decay on the CMB.

Given the deposition history, we can derive constraints on DG dark matter by using the principal-component analysis (PCA) described in, e.g. \cite{Finkbeiner2012,Slatyer2013DarkAges}. We qualitatively describe this method here but refer interested readers to \cite{Finkbeiner2012} for a detailed discussion. Using equations (\ref{decaytophoton}), (\ref{decaytoelectron}), (\ref{eqn:injectionrate}), and (\ref{eqn:depositionrate}), we can write the energy deposition history as 

\begin{align}
    \left( {\frac{d E}{dV dt}} \right)_{{\rm dep},c} = \tilde{f}_c(z) \rho_{\rm DM, 0} (1+z)^3/\tau,
\end{align}

\noindent where $\tilde{f}_c(z)$ is proportional to $f_c(z)$ and $\tau$ is a dimensionful normalization constant (in this case, the lifetime). We can project the redshift dependence of our DG decay energy deposition onto a set of orthonormal basis functions (principal components), $e_i(z)$ derived in \cite{Finkbeiner2012}. Each principal component has an associated error bar, $\sigma_i$, which are in units of $1/\tau$. The $2\sigma$ constraint on $\tau$ can then be written as

\begin{align}
    \tau > \frac12 \left[ \displaystyle\sum_{j = 1}^{N_{\rm pc}} \left( \frac{\tilde{f} \cdot e_j} {\sigma_i} \right)^2 \right]^{1/2},
\end{align}

\noindent where the dot product is defined appropriately, and $N_{\rm pc}$ is the number of principal components being used to derive the constraint. We have defined a total $\tilde{f}$ as the sum of $\tilde{f}_c$ over the following channels: H ionization, He ionization, heating, and excitation. The $2\sigma$ constraints from \emph{Planck} and from an idealized cosmic variance-limited experiment (CVL) are displayed in Fig. \ref{fig:Constraints}.  

We have also computed CMB constraints using the publicly available \texttt{ExoCLASS} code \cite{Stocker:2018avm}. This is a modified version of the Boltzmann solver \texttt{CLASS}~\cite{Blas:2011rf} that is written to include the impact of energy injections on the CMB power spectrum. The main effect of such energy injections is to alter the ionization history of the universe (see also Section~\ref{sec:LyAlpha} below), that is to say the free electron fraction $x_e$ as a function of redshift. This quantity enters in the definition of the Thomson optical depth:
\begin{align}
    \tau(z) = \int_0^z dt n_H(z) x_e(z) \sigma_T
\end{align}
and the visibility function:
\begin{align}
    g(z) \equiv e^{-\tau} \frac{d\tau}{dz}.
\end{align}
These changes can leave important modifications in the CMB power spectrum. This is most easily seen by considering the line-of-sight formalism~\cite{Seljak:1996is}. In this formalism the observed CMB spectrum is obtained by integrating the primordial perturbation spectrum against transfer functions. These latter functions are themselves computed as integrals over the line-of-sight of perturbation source terms that depend on the optical depth and visibility function (and its time derivatives). Changes to the optical depth and visibility functions (even if the perturbations remain unchanged) affect light propagation in cosmic history and leave an imprint on the CMB power spectrum. For a more extensive discussion of energy injections and the effect they leave on the CMB, we refer the reader to~\cite{Poulin:2015pna,Poulin:2016anj}.

In order to cross-check our constraints derived using the PCA method discussed above, we take the simplest approach and provide the energy injection profiles, along with the ``$f$-curves'', as an external input to the code. Interfacing {\texttt ExoCLASS} with \texttt{MontePython}~\cite{Audren:2012wb,Brinckmann:2018cvx}, we run a Markov chain Monte Carlo (MCMC) to determine constraints on the parameter $\lambda$ in the decay rates~\eqref{decaytophoton} and \eqref{decaytoelectron}. This also allows us to also confirm that there are no degeneracies between $\lambda$ and the cosmological parameters as expected from models of decaying DM~\cite{Slatyer:2016qyl}. We find remarkable agreement between the two methods described above. The CMB constraints are displayed in Fig. \ref{fig:Constraints}. 

Our model has two important features that distinguish it from other models of dark matter. The first is the presence of a large number of particles in the dark sector leading to a spread in dark matter masses as shown in Fig.~\ref{fig:MassSpectrum}. Conventional dark matter candidates are instead composed of a single particle with a particular mass. The mass spread has a small effect on the CMB constraints derived from our model. While having particles at lower mass (than the peak) does reduce the decay rate, the mass distribution also has support at masses higher than the peak where the decay rate is larger. These opposing effects result in constraints that are almost unchanged from the case without mass spread. The second feature is the time-evolution of the dark matter mass distribution as shown in the same Figure (and also can be seen from equation~\eqref{eq:massDM}). More conventional dark matter models have a mass that is typically constant in time (although see~\cite{Agrawal:2019dlm} for an example of a model with variable mass dark matter). The time-evolution of the mass distribution also has an impact on the decay rates to SM particles since these have the dependence $\Gamma \propto m^3$. To assess this impact we have to compare our constraints to a model where the DG mass distribution is time-independent. Importantly, in the dark dimension scenario, the dark gravitons start with a mass distribution peaked at $\sim$ GeV where this value is set by demanding that we get the correct DM abundance observed today. The two cases to contrast are then a time-independent DG distribution with peak around $\sim$ GeV and the time-dependent one we study in this paper. For the CMB constraints the strongest effects come from energy injections around $z\sim 600$ \cite{Finkbeiner2012,Poulin:2015pna}. For the time-dependent case, the DGs at $z\sim 600$ have masses in the range of about 1.5 MeV, a factor of 600 smaller than the $\sim$ GeV mass they would have in the time-independent case. A quick estimate then shows that the constraint on $\lambda$ in the time-independent case would be a factor of about $10^4$ stronger and the model with $\lambda \sim \mathcal{O}(1)$ would be ruled out. The time-dependence therefore has a strong impact in alleviating the CMB (and other) constraints.

\subsubsection{Extragalactic Background}

The extragalactic background light (EBL) is the accumulation of radiation emitted throughout the history of the universe. It spans $\sim$ 20 decades of energy from radio to $\gamma$-ray frequencies. Contributions to the EBL come from, e.g., the CMB and emission from stars, galaxies, and active galactic nuclei (AGN). The energies relevant for our analysis lie in the keV--MeV range (i.e. $X$-ray and $\gamma$-ray photons) and we will therefore focus on the cosmic $X$-ray background (CXB) portion of the EBL spectrum. While the origin of the CXB is uncertain, various astrophysical models exist that can explain the observed spectrum in this energy range. These models typically rely on the $X$-ray emission from a population of AGNs as they accrete nearby gas. Models can differ in the assumptions they make on the spectra of these $X$-ray emission, the AGN populations and their surrounding environments. See for instance~\cite{Aird:2015fya,Khaire:2018mvl,Ananna:2018uec} and references therein for examples of models explaining the CXB. 

Apart from astrophysical sources, dark matter decaying outside the Milky Way can contribute to the EBL~\cite{EBL2001}. To compare to observations, we consider the decay of DGs to photons and calculate the differential flux per unit energy, DG mass and solid angle. Here we define flux ($\Phi$) in units of photons cm$^{-2}$ s$^{-1}$. Generalizing the formulae in~\cite{Blanco:2018esa,Ibarra:2007wg,Chen:2009uq} for example, we immediately find: 
\begin{align}
    E^2\frac{d\Phi}{dE d\Omega dm} =  \frac{E^2}{4\pi} \int &\frac{dz}{H(z)(1+z)^3} \times \nonumber \\
    &\Gamma_{{\rm KK}, \gamma} \frac{\rho_c(z)}{m}\frac{d\ln\rho}{dm}(z,m) \frac{dN}{dE'}(E' = E(1+z))e ^{-\kappa(z,E'=E(1+z))}
    \label{eqn:dphidedm}
\end{align}
where $dN/dE = 2\delta(E - m/2)$ is the spectrum produced by the decay of a single KK graviton of mass $m$ and
\begin{align} \label{eqn:Hion}
    \kappa(z,E) &= \displaystyle\int_0^z {\frac{dz'} {(1+z') H(z')}} n_H(z') \sigma_{pe}(E),
\end{align}

\noindent is the attenuation coefficient due to Hydrogen photoionization \cite{Cadamuro2012}. In (\ref{eqn:Hion}), $n_H(z)$ is the number density of atomic hydrogen, $\sigma_{pe}(E) = 256\pi(E_{1s}/E)^{7/2}/(3\alpha m_e^2)$ is the photoelectric cross-section and $E_{1s} = 13.6$ eV is the ionization energy of Hydrogen. Integrating (\ref{eqn:dphidedm}) over KK mass states we get the flux

\begin{align}
E^2 \frac{d\Phi}{dE d\Omega} = &\frac{E}{2\pi} \frac{\rho_{\mathrm{crit},0}\Omega_c}{H_0} \int_{2E}^\infty \frac{dm}{\sqrt{\Omega_\Lambda +\Omega_c (m/2E)^3}} \times \nonumber \\
& \frac{\Gamma_\gamma(m)}{m}
e^{-\kappa(-1 + m/2E, m/2)} \frac{d\ln\rho}{dm}\left(z = -1 + \frac{m}{2E},m\right).
\end{align}

\begin{figure}
    \centering
    \includegraphics[width=\linewidth]{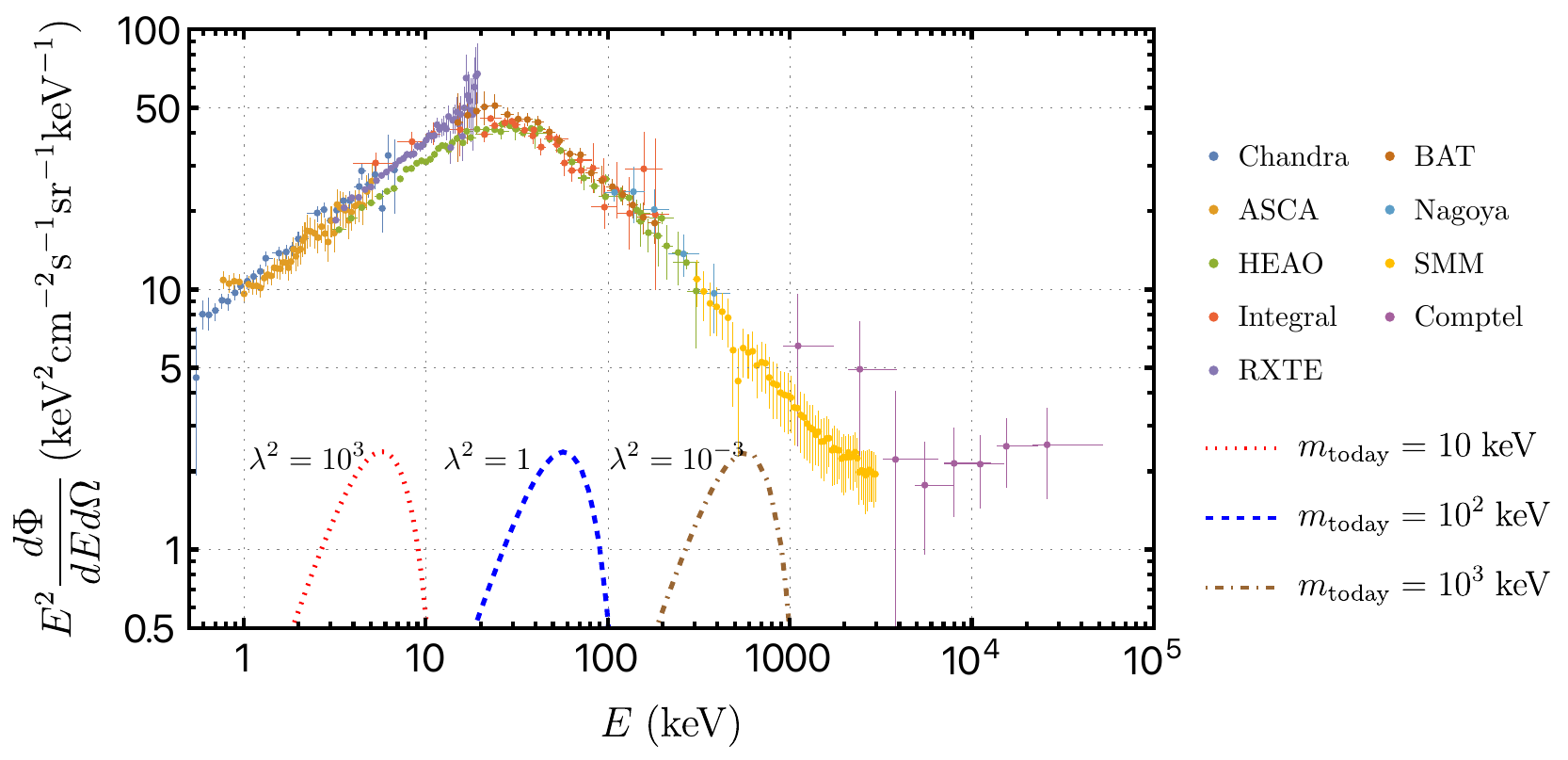}
    \caption{We show the $X$-ray and $\gamma$-ray differential flux measurements from Chandra~\cite{Cappelluti:2017miu}, ASCA~\cite{gendreau1995asca}, HEAO~\cite{Gruber:1999yr,kinzer1997diffuse}, Integral~\cite{churazov2007integral}, RXTE~\cite{Revnivtsev:2003wm}, {\it Swift}/BAT~\cite{Ajello:2008xb}, Nagoya~\cite{fukada1975energy}, SMM~\cite{watanabe1999diffuse} and Comptel~\cite{weidenspointner2000cosmic}. The curves show the expected signal from the decay of dark gravitons to photon for a few representative masses. The normalization of the curves is arbitrary and it scales with $\lambda^2$ as in Eq.~\eqref{decaytophoton}.}
    \label{fig:CXB}
\end{figure}

An example of this contribution for some representative DG central masses today and coupling $\lambda$ is shown in Figure~\ref{fig:CXB} where we overlay current CXB observations from various missions. To derive constraints on KK gravitons we use EBL data in the energy range close to the peak of the DG expected spectrum. For example, for DG with $m_\mathrm{today} = 100$ keV, the spectrum peaks at an energy of about $E \approx 50\;\mathrm{keV}$. In this instance, we use the data from the HEAO, Integral, RXTE and {\it Swift}/BAT missions~\cite{Cappelluti:2017miu,gendreau1995asca,Gruber:1999yr,kinzer1997diffuse,churazov2007integral,Revnivtsev:2003wm,Ajello:2008xb,fukada1975energy,watanabe1999diffuse,weidenspointner2000cosmic}. For other dark matter masses, we use the relevant subset of datasets from Figure~\ref{fig:CXB}. 

After determining the relevant datasets, we derive two separate constraints on the decay rate of DGs to photons. The first is a conservative constraint where we simply require that the signal from DG emission is smaller than the observed EBL flux. This approach leads to weaker bounds but has the advantage that it is agnostic to uncertainties in astrophysical modeling. We also derive a more aggressive bound on the decay rate parameter $\lambda$ by assuming that astrophysics can explain the full EBL spectrum. In this case, we require that the contribution from DG decays be smaller than the measurement uncertainties. In practise, we assume that all data points are at zero flux with the same error bar and find the 2$\sigma$ exclusion limit on $\lambda$. Both of these constraints are shown in Figure \ref{fig:Constraints}.

The constraints are derived by defining a simple $\chi^2$ likelihood and sampling from the posterior of $\lambda$ using the Metropolis-Hastings algorithm. The distribution of $\lambda$ samples is then proportional to the posterior and we use this to obtain the 95\% confidence intervals for the coupling $\lambda$. 

Finally, we discuss the importance of the DG mass profile and its time-dependence on EBL constraints. The mass profile has very little effect on the shape and amplitude of the flux signal. We mention that there may be direct observational strategies that are more sensitive to the mass profile in the dark dimension model. For instance searching, for peaks in galactic emission spectra with a shape implied by the distribution in Figure \ref{fig:MassSpectrum}. These searches can target DM rich galaxies where the signal would be stronger and easier to distinguish from background emission (see for example~\cite{Grin:2006aw}). We now turn to the importance of the time-dependence. As in the case of the CMB, the time-dependence plays a crucial role. In fact the contribution to the signal in figure~\ref{fig:CXB} comes mostly from low redshift $z\lesssim 1$ where the DG mass profile is peaked very near to $m_\mathrm{today}$. In the case where $m_\mathrm{today} \sim 100$ keV, this is a factor of about $10^4$ smaller than the typical $1 \sim$ GeV masses of DGs at production. The constraint on $\lambda$ with the time-dependence is therefore weaker by a factor of about $10^6$ compared to the case with no time-dependence. 

\subsubsection{Lyman $\alpha$}
\label{sec:LyAlpha}
Recently Lyman-$\alpha$ data, with the aid of simulations, allowed for a measurement of the IGM temperature at low redshifts~\cite{Walther:2018pnn,Gaikwad:2020art}. If DM decays and injects energy into the IGM, it could raise the IGM temperature to higher values than the measurements. These temperature measurements therefore place upper bounds on the amount of energy injection into the IGM.

To derive these constraints, we use a modified version of the public code \texttt{DarkHistory} \cite{Liu:2019bbm,Liu:2020wqz, liu2023exotic, liu2023exotic2}. Among other things, this code keeps track of two coupled quantities, the ionization fraction $x_e$ and the matter temperature $T_m$, describing the IGM and self-consistently solves for their evolution. These quantities obey a system of coupled differential equations given by:
\begin{align}
    \dot{x}_{\mathrm{HII}}&=\dot{x}_{\mathrm{HII}}^{\text {atom }}+\dot{x}_{\mathrm{HII}}^{\mathrm{DM}}+\dot{x}_{\mathrm{HII}}^{\star} \\ 
    \dot{T}_{\mathrm{m}}&=\dot{T}_{\mathrm{adia}}+\dot{T}_{\mathrm{C}}+\dot{T}_{\mathrm{DM}}+\dot{T}_{\mathrm{atom}}+\dot{T}^{\star}.
\end{align}
In the first line the terms correspond to ionization due to atomic processes, DM energy injection and astrophysical energy injections (e.g. from stars). In the second line the terms correspond to adiabatic cooling, Compton cooling/heating, heating due to energy injection, atomic cooling and stellar heating. More information on these quantities is given in the supplemental materials of~\cite{Liu:2019bbm}. This signature is related to the one used to derive constraints on $\lambda$ from the CMB (see section~\ref{sec:CMB}) although in the case of the CMB, it is the ionization fraction $x_e$ that plays an important role.

We will follow the conservative approach taken in~\cite{Liu:2020wqz} where we set the astrophysical source terms to zero and place only an upper bound on the IGM temperature (i.e. allow models with lower IGM temperature than the measurement, see below). This has the advantage of not relying on models of astrophysical sources which could have large uncertainties (see for example~\cite{McQuinn:2015icp}). In this conservative approach, a model of the IGM temperature is only penalized if the temperature exceeds the observed value. A model that produces a lower IGM temperature is not penalized since it can potentially be made consistent with observations by the addition of astrophysical sources that are ignored in the conservative limit. For the ionization history, we use a Tanh profile with bounds derived from {\it Planck} measurements. We derive bounds for two fiducial reionization models, an early and a late Tanh profile, consistent with {\it Planck} measurements errors. Marginalizing over ionization histories with more general functional forms typically has only a small effect on the constraints derived on model parameters (as in~\cite{Liu:2020wqz}). The small difference we find between the early and late Tanh models also supports this conclusion.

In order to calculate model constraints from \texttt{DarkHistory}, we need to determine the particle production rate as a function of redshift which is different in our model compared to the usual decaying DM. We consider the event rate for producing particles of energy $E$ per mass bin:
\begin{align}
    \frac{dN_\mathrm{events}}{dV dt dE dm} &= \left[\Gamma_\gamma  \delta\left(E-\frac{m}{2}\right) + \Gamma_e \delta\left(E - \frac{m-2m_e}{2}\right)\right] \frac{d\ln\rho}{dm}(z,m) \frac{\rho_c}{m}.
\end{align}
This can be integrated over masses and/or energies to give the following information required to calculate the impact of our model on the IGM temperature as the universe evolves:
\begin{itemize}
    \item The energy spectrum of injected photons and electrons/positrons:
    \begin{align}
        \frac{dN_\mathrm{events}}{dV dt dE} = \int dm\; \frac{dN_\mathrm{events}}{dV dt dE dm}
    \end{align}
    \item The total number of decay events per unit volume per unit time:
    \begin{align}
        \left(\frac{dN_\mathrm{events}}{dV dt}\right) = \int dE dm\; \frac{dN_\mathrm{events}}{dV dt dE dm}
    \end{align}
    \item The total energy injection rate per unit volume per unit time:
    \begin{align}
        \left(\frac{dE}{dV dt}\right) = \int dE dm\; m \frac{dN_\mathrm{events}}{dV dt dE dm}
    \end{align}
\end{itemize}

Our energy injection can then be described using the same formalism outlined in~\cite{Liu:2019bbm}. For example, the spectrum of injected photons or electrons/positrons at a particular redshift $\overline{S}^\alpha$ with $\alpha \in\{\gamma, e\}$ can be deduced from the relation:
\begin{align}
    \frac{dN^\alpha}{dE} = \overline{S}^\alpha \left(\frac{dN}{dV dt}\right) G(z)
\end{align}
where 
\begin{align}
    G(z) = \frac{\Delta \log (1+z)}{n_B(z) H(z)} = \frac{\Delta t}{n_B(z)}
\end{align}
converts an event rate per volume into the number of events per baryon\footnote{Here, as in~\cite{Liu:2020wqz}, the energy is the kinetic energy of the injected particle.}.

Given the event and energy injection rates as well as the spectra of injected photons and electrons, we leverage the machinery developed in~\cite{Liu:2019bbm,Liu:2020wqz} and made public in the {\texttt DarkHistory} code to solve for the evolution of ionization fraction and IGM temperatures. Comparing the resulting solution with the data using the conservative approach outlined above, we derive the Lyman-$\alpha$ limits shown in figure~\ref{fig:Constraints}. 

We now turn to a discussion of how the features of our model impact the constraints derived from Lyman-$\alpha$ temperature data. The major difference with conventional dark matter candidates lies in the time-dependence of the mass. The time-dependence of the DM mass profile leads to weaker Lyman-$\alpha$ constraints compared to the time-independent case. This can be seen, for example, by extrapolating the constraint profile in figure~\ref{fig:Constraints} to $m_\mathrm{today} \sim 1$ GeV. As in the case of the CMB, the spread of the DM mass distribution plays a small role. 

\subsection{Late Universe Constraints}

\subsubsection{Milky Way Decay}

Dark dimension gravitons decaying through $KK \to \gamma\gamma$ within the Milky Way contribute to the anisotropic $X$-ray emission observed by numerous telescopes including XMM-Newton, INTEGRAL, NuSTAR and Suzaku. The differential flux, $\Phi$,  from a region of of the sky with galactic coordinates $(\ell, b)$ \footnote{$\ell$ and $b$ are called the `galactic longitude' and `galactic latitude', respectively.} is 

\begin{align} \label{eqn:MWFlux}
    E^2 {\frac{d \Phi} {d E d\Omega}} = {\frac{E}{4\pi}} \Gamma_{{\rm KK}, \lambda}(2E) {\frac{d \ln \rho}{dm}}(2E, z=0) \displaystyle\int ds \rho_{\rm DM}(r(s,\ell,b)),
\end{align}

\noindent where the expression for $\Gamma_{{\rm KK}, \lambda}$ is given in (\ref{decaytophoton}). In the equation above the line-of-sight coordinate is denoted by $s$ and the Milky Way dark matter density by $\rho_{\rm DM}$. For this analysis, we assume a Navarro-Frenk-White (NFW) profile,

\begin{align}
    \rho_{\rm DM}(r(s, \ell, b)) &= \frac{\rho_s}{\frac{r}{R_s}\left(1 + \frac{r}{R_s} \right)^2}, \quad r = \sqrt{s^2 + R^2 - 2 s R \cos \ell \cos b},
\end{align}

\noindent where $\rho_s = 0.18$ GeV$/$cm$^{3}$, $R_s = 24.4$ kpc, and $R = 8.2$ kpc is the distance from the solar neighborhood to the galactic center. The predicted flux from KK graviton decay (\ref{eqn:MWFlux}), can be compared to data from $X$-ray observations in the relevant frequency range, which predominantly come from the INTEGRAL and NuSTAR missions.

The INTEGRAL data collected by the Spectrometer on INTEGRAL (SPI) over 16 years was reported in \cite{Bouchet2005, Bouchet2008, Bouchet2011}. The observed energy range spanned from 27 keV to 1.8 MeV, which is split into the following bins, $b_1 \in [27, 49]$ keV, $b_2 \in [49, 90]$ keV, $b_3 \in [100, 200]$ keV, $b_4 \in [200, 600]$ keV, $b_5 \in [600, 1800]$ keV. As in \cite{Cirelli2021,Pinetti2023}, we only consider latitude bins that exclude the galactic plane and restrict to longitudes with $|\ell| < 23.1^\circ$ for energy bins $b_1-b_4$ and $|\ell| < 60^\circ$ for $b_5$ (see Fig. 4 in \cite{Cirelli2021}).  

To derive constraints we use the test statistic used in \cite{Cirelli2021, Pinetti2023}, 

\begin{align} \label{eqn:chisquare}
    \chi^2_> = \displaystyle\sum_{i=1}^5 \displaystyle\sum_{j=1}^{N_i} \left( { \frac{\max\left[\Phi_{i,j}(\lambda, m_{\rm today}) - \phi_{i,j}, 0 \right]}{\sigma_{i,j}} }\right)^2, 
\end{align}

\noindent where the $i$ runs over the energy bins and $j$ runs over the latitude bins observed in each energy bin. For $i\in{1,2,3,4}$, $N_i = 18$ and $N_5 = 12$. The observed flux values are represented by $\phi_{i,j}$ and the corresponding error on the measurement by $\sigma_{i,j}$. The predicted flux values due to KK graviton decay with peak mass $m_{\rm today}$ and coupling $\lambda$ are represented by $\Phi_{i,j}(\lambda, m_{\rm today})$. We compute $\Phi_{i,j}(\lambda, m_{\rm today})$ by integrating (\ref{eqn:MWFlux}) as follows,

\begin{align}
    \Phi_{i,j}(\lambda, m_{\rm today}) = \displaystyle\int_{E_{{\rm min}, i}}^{E_{{\rm max},i}} dE \displaystyle\int_{\ell_{{\rm min}, i}}^{\ell_{{\rm max}, i}} d\ell  \displaystyle\int_{b_{{\rm min}, i,j}}^{b_{{\rm max},i,j}} db \ \cos(b) {\frac{d \Phi}{ dE d\Omega}},
\end{align}

\noindent where $E_{{\rm min (max)},i}$ is minimum (maximum) energy in bin $i$, $\ell_{{\rm min (max)},i}$ is the minimum (maximum) longitudes considered in bin $i$, and $b_{{\rm min (max)},i,j}$ is minimum (maximum) latitude of latitude bin $j$ in energy bin $i$. In Fig. \ref{fig:Constraints}, we show $2\sigma$ constraints derived by setting $\chi^2_> = 4$. We note that these constraints are conservative as we have not subtracted from $\phi_{i,j}$ any known astrophysical backgrounds. 

The Nuclear Spectroscopic Telescope Array (NuSTAR)\footnote{The data were collected from two separate detectors aboard NuSTAR, FPMA and FPMB. The spectra collected by the two detectors are very similar and thus lead to almost identical constraints. Therefore, we only derive constraints from FPMA.} has made sensitive measurements of $X$-ray emission in blank-sky regions~\cite{Perez2017, NuSTARbso2021, Roach2023}, the galactic center (GC)~\cite{Mori_2015, Hong_2016}, and faint-sky regions directly above and below the galactic plane~\cite{Roach2020}. For blank and faint-sky observations the dominant astrophysical background is the cosmic $X$-ray background (CXB). We discussed constraints from CXB observations previously and focus here on constraints from GC emission. The data used to derive the constraint is available in Fig. 5 of \cite{Perez2017}. Following \cite{Pinetti2023}, we approximate the GC emission region as an annulus around the GC with inner radius $1.5^\circ$ and outer radius $3.5^\circ$. In \cite{Perez2017} the data is binned logarithmically with 200 equally spaced bins per decade of energy. We compute the photon flux from DG decay in the emission region, $\Phi_i$, where $i$ runs over energy bins, and compare to the data. To compute the sensitivity we follow the same procedure described above. The constraint in Fig. \ref{fig:Constraints} is the 2$\sigma$ constraint derived by setting $\chi^2_> = 4$. 

The shaded region in Fig. \ref{fig:Constraints} is conservative and can be improved in a number of ways. The data presented in \cite{Perez2017} receive contributions from known astrophysical sources including the CXB and galactic ridge $X$-ray emission (GRXE). Emission from the GRXE is thought to be made up predominantly of discrete sources such as accreting white dwarfs with active coronae \cite{GRXE2009}. Contributions from the CXB and GRXE have been modeled in \cite{Perez2017} and can be subtracted from the observed data to achieve the more stringent constraints on DG decay shown in Fig. \ref{fig:Constraints}. We emphasize that these constraints are less robust as they are sensitive to astrophysical modeling uncertainties. We also comment that the binning procedure utilized in \cite{Perez2017} was optimized to search for line-like emission from sterile neutrino decay. The bin width is much smaller than the typical width of the DG mass distribution, leading to a suppression in the sensitivity to DG decay. The suppression due to the DG mass spread compared to a monochromatic mass distribution is a factor of $\sim \sqrt{\delta E/E} \approx 0.22$ in $\lambda$, where $\delta E/E \approx 0.05$ is the bin width used in \cite{Perez2017}. 
   
\subsubsection{511 keV Line}

The SPI aboard the INTEGRAL satellite has observed line emission at 511 keV coming from the Galactic bulge and disk \cite{511keV2005, Prantzos2011}. Emission from the GC has been observed with a flux of $\sim 10^{-3} \ {\rm cm}^{-1} \ {\rm s}^{-1}$. The observed flux is well-fit by a three-component spectrum made up of narrow and broad ($5.4\pm 1.2$ keV FWHM) features and continuum flux at $E_\gamma < 511$ keV. The observed gamma ray flux implies a positron production rate of $\sim 2 \times 10^{43} \ e^+ \ {\rm s}^{-1}$ in the galaxy. Positrons are believed to be primarily produced as secondaries during interactions of high energy cosmic rays with the interstellar medium. Possible sources of galactic cosmic rays include supernovae and supernova remnants, pulsars (including millisecond pulsars and magnetars)~\cite{HardingLai2006}, $X$-ray binaries~\cite{Grimm2002}, and the central black hole, Sag A$^*$~\cite{Riegler1981}. 

It has also been suggested that the excess can be explained by various dark matter models including annihilating MeV-scale dark matter \cite{Boehm2004, Huh2008, Hooper2008, Khalil2008}, decaying supersymmetric dark matter \cite{Hooper2004, Cembranos2008, craig2009modulino}, upscattered dark matter \cite{Finkbeiner2007, Pospelov2007, Cline2011}, and evaporating primordial black holes \cite{FRAMPTON2005, Bambi2008, Dasgupta:2019cae, Laha2019, DeRocco2019, Cai2021, Keith2021}. Given the morphology of the signal, it has been argued that the 511 keV excess cannot be due to decaying dark matter as the flux would then track the DM density in the galaxy contradicting observations which find a more peaked flux profile (see for example~\cite{Vincent:2012an}). As such, our interest in the 511 keV signal is simply to provide a constraint on our model parameters from requiring the DG decays not produce too many positrons.

Following \cite{DeRocco2019}, we set conservative constraints on decaying DGs by requiring that the galactic positron injection rate not exceed $dN_{e^+}/dt = 4 \times 10^{43} \ {\rm s}^{-1}$, twice the value required to explain the flux from the GC. Importantly, this constraint is agnostic to astrophysical sources of positrons and robust against modeling uncertainties. The production rate of positrons with energies between $E_{\rm min} = m_e$ and $E_{\rm max} = 1$ MeV is 

\begin{align} \label{eqn:positronrate}
    {\frac{d N_{e^+}} {dt}} = \displaystyle\int_{2 E_{\rm min}} ^{2 E_{\rm max}} dm \displaystyle\int d^3 r {\frac{\rho_{\rm DM}(r)} {m}} {\frac{d \ln \rho}{dm}} \Gamma_{{\rm KK}, e^{\pm}} (m).
\end{align} 

The upper boundary on the positron energy, $E_{\rm max}$, is set by the requirement that produced positrons must not escape the halo before annihilating \cite{Alexis2014}. 

As mentioned above, the bounds can be improved markedly with better understanding of astrophysical foregrounds. In addition to sources of positrons, there is considerable uncertainty in the propagation of  $\sim$ MeV positrons in the magnetized, turbulent interstellar medium. This affects the maximum positron energy ($E_{\rm max}$) that goes into (\ref{eqn:positronrate}). Finally, the spatial morphology of decaying DM does not match the observed morphology of the signal. Taking this into account would strengthen the constraints. For our model, constraints from 511 keV set in for $m_{\rm today} \gtrsim m_e$, where CMB constraints rule out $\lambda \gtrsim 0.01$. Thus, we do not pursue the aforementioned avenues for improving 511 keV bounds.

\begin{figure}
    \centering
    \includegraphics[width=0.80\linewidth]{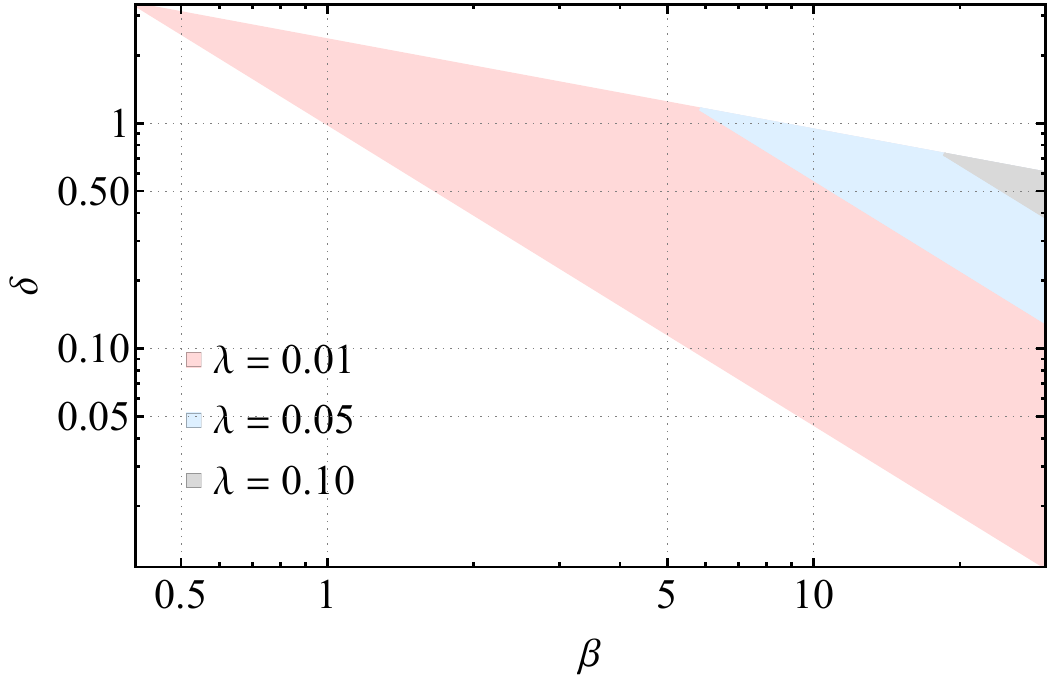}
    \caption{Constraints in the $(\beta,\delta)$-plane for a few representative values of $\lambda$ and for $\gamma = 1$. The triangles show the allowed values for parameters $(\beta, \delta)$ for each $\lambda$. When a pair $(\beta, \delta)$ is not excluded for one value of $\lambda$ then it is also not excluded for any lower $\lambda$ (i.e. the region for $\lambda = 0.05$ includes that of $\lambda = 0.1$ for example). }
    \label{fig:betadeltaPlot}
\end{figure}

\section{Results and conclusion}
\label{sec:results}
We have seen that various astrophysical probes constrain the dark matter mass and the parameter $\lambda$ in the dark dimension scenario. However, as outlined in~\cite{Gonzalo:2022jac}, there could be other $\mathcal{O}(1)$ numbers parameterizing our ignorance of the higher dimensional theory.  In this section, we will interpret the results we obtained from experimental constraints in terms of these parameters. In particular, we will show an exciting interplay between the astrophysical constraints derived in this paper and other constraints from the literature.

Let us begin by reviewing the important parameters for phenomenology (some of these were also mentioned briefly in Section~\ref{sec:DDPheno}):
\begin{itemize}
    \item $\lambda$: this is the coupling of SM fields to the bulk graviton, which is sensitive to the wave function of the KK modes at the SM brane.  It in particular controls the coupling between the dark sector and the visible sector. This influences the DG decays to the SM and production rates from the SM.
    \item $\beta$: The 
    $\beta$ coefficient controls the coupling to other dark gravitons in the tower and affects the decay rate within the DG tower and correlates with the amplitude of inhomogeneities in the 5-th dimension.
    \item $\delta$: this is a measure of the wavelength of 5th dimension inhomogeneities and determines the allowed KK momentum violation in each DG decay. 
    \item $\gamma$: this is the ratio of the mass of the lightest KK graviton to $m_{KK}$ which is the graviton mass spacing at asymptotically large KK quantum numbers. This is introduced to parameterize the constraints of fifth force experiments since these are sensitive to the mass of the first graviton mode rather than $m_{KK}$ (see also equation~\eqref{eq:KKbound} below).
\end{itemize}

Before starting the main discussion, we comment on a relation between the KK scale and the value of $\lambda$ implied by the DG production mechanism proposed in~\cite{Gonzalo:2022jac}. Recall that the DGs are produced from the SM brane via gravitational strength interactions. The abundance of DGs is determined by the model parameters and the initial temperature which can be chosen to produce the observed amount of the DM. The only restriction on the initial temperature is that it must be higher than the BBN temperature $T_{BBN} \sim 1\;\mathrm{MeV}$. This condition gives the following relation between the KK scale and the coupling $\lambda$:
\begin{align}
    \lambda < 2\times 10^4 \left(\frac{m_{KK}}{\mathrm{meV}}\right)^{1/2} \qquad \text{(from BBN)}.
\end{align}
This bound is much weaker than the bounds we derived in this paper and we do not discuss it further. 

Let us now move on to the discussion of the model parameters. We will start by reviewing two relations that will be important in what follows. As we saw already the DGs continually decay to lighter dark gravitons as the universe evolves. The decay rate is given in~\eqref{eq:totWidth} and leads to the time dependence shown in~\eqref{eq:massDM}. Since the distribution of DGs is peaked around the mass scale whose lifetime is Hubble $H \sim 1/t$, we can in fact obtain the following expression from~\eqref{eq:totWidth} for the mass:
\begin{align}
    \label{eq:massDMwithParams}
    m_{DM}(t) \approx \frac12 \left(
    \frac{M_P^4 m_{KK}}{\delta^3 \beta^4}
    \right)^{1/7}\frac{1}{t^{2/7}}
\end{align}
where we included an approximate factor of $\frac12$ to better match the peak position found in numerical simulations. One can see that larger values of $\beta$ and $\delta$ lead to smaller DG masses as they mean faster decays within the tower. This is the first expression we will use.

The second equation expresses the velocity of the daughter particles (i.e. the kick velocity) due to DG decay:
\begin{align}
        \label{eq:DGvelocity}
    v \approx \sqrt{\frac{\delta m_{KK}}{m_{DM}}}.
\end{align}
This relation is easy to obtain by equating the kinetic energy of the two daughter particles $m_{DM} v^2$ to the total mass loss in the decay $\delta m_{KK}$. Here we see that higher values of $\delta$ mean larger kick velocities as the decay can convert more of the parent DG mass to kinetic energy. The model is therefore parameterized by 5 numbers $(\delta, \beta, \lambda, \gamma, m_{KK})$.

On the other hand, we have experimental bounds on various quantities that enter in the latter two expressions above. The first constraint comes from lack of observation of deviations from Newton's law at short distances~\cite{Tan:2016vwu,Lee:2020zjt}. This implies a lower bound on the KK scale:
\begin{align}
    \label{eq:KKbound}
    m_1 \equiv \gamma m_{KK} > 6.6 \;\mathrm{meV} \qquad \text{(fifth force experiments)}.
\end{align}
\noindent The second experimental constraint can be obtained by studying the cosmology of the DGs and how they affect structure formation. In particular, the kick velocity in~\eqref{eq:DGvelocity} can suppress structure formation at small scales and can be in conflict with observation if it is too large. In an upcoming paper~\cite{UpcomingPaper}, this is studied and it is found that:
\begin{align}
    \label{eq:velocityUpperBound}
    v \approx \sqrt{\frac{\delta m_{KK}}{m_{DM}}} < 2.2 \times 10^{-4} \qquad \text{(from linear cosmology)}.
\end{align}
Finally, we have the astrophysical bounds derived in this paper~\eqref{eq:CMBBound}. We will rewrite this in the form:
\begin{align}
    \label{eq:AstroBound}
    m_{DM} < \left(\frac{0.1}{\lambda}\right)^{2/3} \times 100 \;\mathrm{keV} \qquad \text{(this work)}.
\end{align}
These are 3 inequalities that cut off certain regions of the 5-dimensional parameter space $(\delta, \beta, \lambda, \gamma, m_{KK})$. We will now derive various constraints implied by the above inequalities to get intuition for this allowed region in parameter space. 

\begin{figure}
    \centering
    \includegraphics[width=0.80\linewidth]{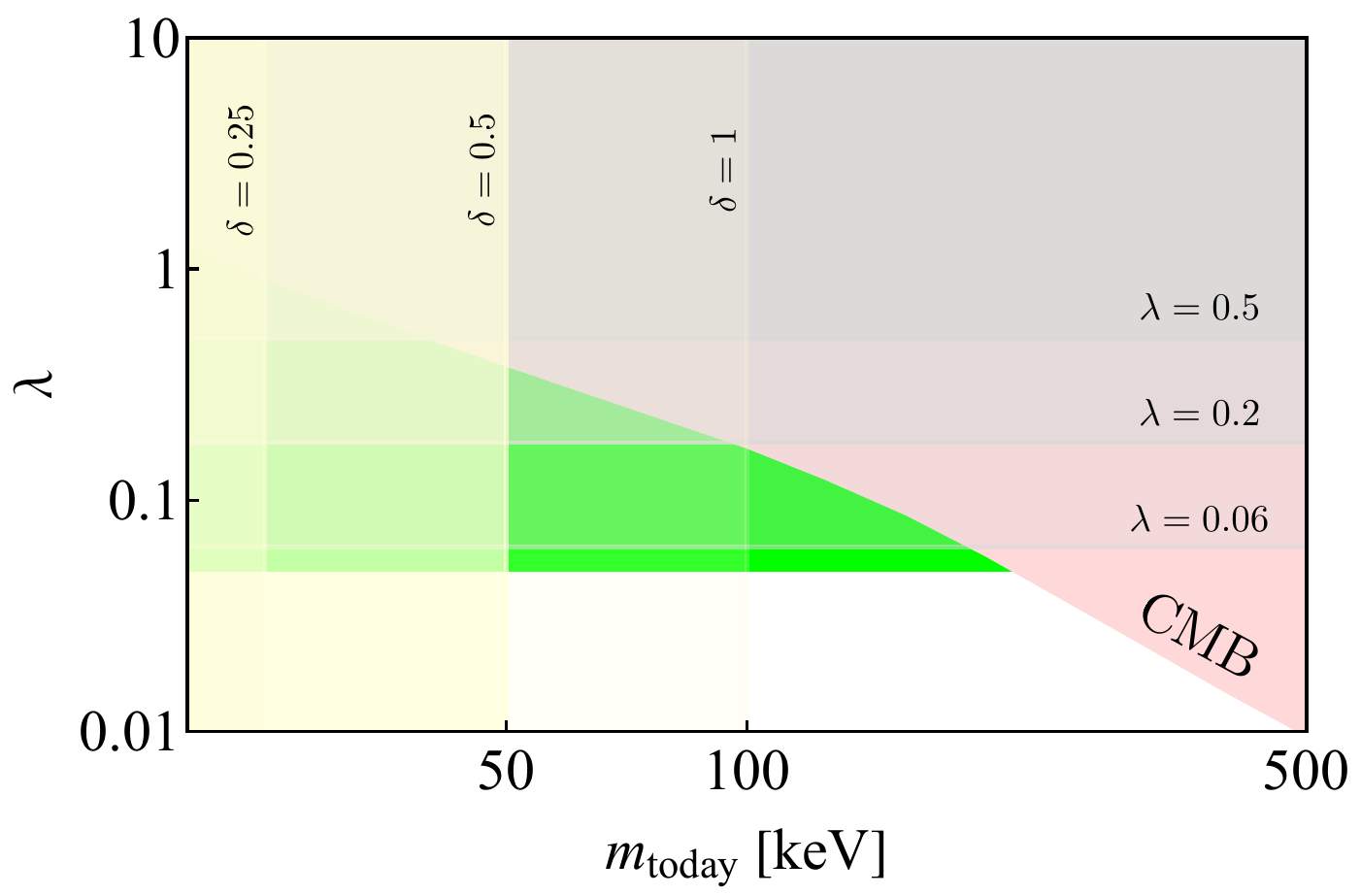}
    \caption{The natural mass and $\lambda$ coupling range in the dark dimension cosmological scenario for the parameter $\gamma = 1$. The red region is the exclusion bound from the CMB derived in this work. The yellow vertical regions show the bound in~\eqref{eq:mDMBound} for a few values of $\delta$. The grey horizontal regions show the corresponding bounds for $\lambda$ from~\eqref{eq:deltaLambdaBound}. For example, when $\delta = 1$, we have $\lambda < 0.06$ (for $\gamma = 1$). The filled green region shows the mass range and coupling $\lambda$ with $\mathcal{O}(1)$ parameters. }
    \label{fig:allowedRegion}
\end{figure}

First, we show that the above inequalities imply a lower bound on $m_{DM}$. This is simple to do by using the upper bound on the velocity $v$ given in~\eqref{eq:velocityUpperBound} and the lower bound on $m_{KK}$ in~\eqref{eq:KKbound}. This gives:
\begin{align}
    \label{eq:mDMBound}
    m_{DM} > \frac{\delta}{\gamma} \times 100 \;\mathrm{keV}.
\end{align}
This area is shown by the yellow vertical regions in Figure~\ref{fig:allowedRegion} for $\delta \in  \{0.5, 1, 2\}$ and $\gamma = 1$.

Next, using the lower bound on $m_{KK}$ from~\eqref{eq:KKbound} and the lower bound on $m_{DM}$ from~\eqref{eq:AstroBound} in the inequality~\eqref{eq:velocityUpperBound}, we immediately find:
\begin{align}
    \label{eq:deltaLambdaBound}
    \delta < 0.2 \times \gamma \lambda^{-2/3}.
\end{align}
We take this opportunity to comment on the value of $\delta$ which need not be strictly integer. As mentioned previously, $\delta$ captures the smoothness of the dark dimension and induces couplings between KK gravitons with masses that do not obey KK momentum conservation (i.e. the KK momenta do not add up to zero). However, in general, there might different couplings amongst different KK modes with different KK violation for each of these couplings. Our parameter $\delta$ should be thought of as a suitable average of these violations and therefore should not be restricted to take integer values.

We can also use the expression for $m_{DM}$ in inequality~\eqref{eq:AstroBound} and the lower bounds on $m_{KK}$ and $\delta^{-3}$ to arrive at an inequality involving only $\beta$, $\gamma$, and $\lambda$. This gives:
\begin{align}
    \label{eq:betaLambdaBound}
    \beta > 800 \times \gamma^{-1} \lambda^{5/3}.
\end{align}

In addition to the above two inequalities, we can analyse the constraints differently to obtain limits in the $(\beta, \delta)$-plane. First, one can isolate $\delta^3 \beta^4$ in the expression for $m_{DM}$ and use the lower bounds on $m_{KK}$ and $m_{DM}$ from~\eqref{eq:KKbound} and~\eqref{eq:AstroBound} to get:
\begin{align}
    \label{eq:betadeltaLowerBound}
    \beta^2 \delta^{3/2} > \gamma^{-1/2}(100\lambda)^{7/3}.
\end{align}
There is also an upper bound on $\beta^2 \delta^5$ that can be obtained directly using equation~\eqref{eq:mDMBound}:
\begin{align}
    \label{eq:betadeltaUpperBound}
    \beta^{2/3} \delta^{5/3} < 4\gamma
\end{align}
Given a value of $\gamma$ and $\lambda$ consistent with eq.~\eqref{eq:deltaLambdaBound}, the above two inequalities show the region in the $(\beta, \delta)$-plane that is allowed by current observations. We show this region for a few representative values of $\lambda$ and $\gamma = 1$ in Figure~\ref{fig:betadeltaPlot}. Note that these two inequalities imply~\eqref{eq:deltaLambdaBound} and~\eqref{eq:betadeltaLowerBound}. 

Remarkably, the above data also imply an upper bound on the KK scale in terms of the parameters $\delta$ and $\lambda$. This can easily be obtained using the upper bounds on the velocity~\eqref{eq:velocityUpperBound} and the DG mass~\eqref{eq:AstroBound}. Combining this with the bound from fifth force experiments, we have:
\begin{align}
    \label{eq:mKKUpperBound}
     \frac{1}{\gamma} \times 6.6\;\mathrm{meV} < \;& m_{KK} < \left(\frac{0.5}{\delta}\right)\left(\frac{0.1}{\lambda}\right)^{2/3} \times 10\;\mathrm{meV} \\
     \gamma \times 30\;\mu\mathrm{m} > \;& l_{5} > \left(\frac{\delta}{0.5}\right) \left(\frac{\lambda}{0.1}\right)^{2/3} \times 20\; \mu\mathrm{m}\\
          30\;\mu\mathrm{m} > \;& l^{eff}_{5} > \left(\frac{2}{\gamma} \right) \left(\frac{\delta}{0.5}\right) \left(\frac{\lambda}{0.1}\right)^{2/3} \times 10\; \mu\mathrm{m}
\end{align}
where $l^{eff}_5=\gamma^{-1}l_5$ is the effective radius that would be meausured in inverse square law experiments.
In particular, for natural parameter values for our model (say taking $\gamma \sim 2, \delta \sim 0.5, \lambda \sim 0.1$), this large extra dimension is within the sensitivity of upcoming fifth force experiments;  roughly $\mathcal{O}(1)$ parameters would lead to a range for the effective radius of $l^{eff}_5\sim 1 - 30\; \mu$m. We show the natural range of DG mass and coupling in Figure~\ref{fig:allowedRegion}. It is also possible to probe the KK scale by astrophysical means~\cite{Hannestad:2003yd}. In this arena, constraints are obtained from limits on neutron star cooling and strong bounds can potentially be obtained by finding particularly cool neutron stars. On the cosmological frontier, upcoming data~\cite{Troja:2022vmg} will improve our determination of the upper bound on the average kick velocity~\eqref{eq:velocityUpperBound}. In addition, there may be effects in the non-linear regime associated to a fraction of DM having a considerable velocity (see for example~\cite{Dienes:2020bmn}). 

\acknowledgments
We are grateful to Tracy Slatyer, Hongwan Liu, Wenzer Qin, and Yitian Sun for helpful discussion on computing CMB constraints and Elena Pinetti for guidance on deriving $X$-ray constraints. We have also greatly benefited from discussions with Eduardo Gonzalo.

The work of CV is supported by a grant from the Simons Foundation (602883,CV), the DellaPietra Foundation, and by the NSF grant PHY-2013858. AP acknowledges support from the Princeton Center for Theoretical Science. The work of GO work is supported by a Leverhulme Trust International Professorship grant number LIP-202-014. For the purpose of Open Access, the author has applied a CC BY public copyright licence to any Author Accepted Manuscript version arising from this submission. In addition, GO and CV are grateful for the support and hospitality of the Simons Center for Geometry and Physics, Stony Brook University at which some of the research for this paper was performed during the summer workshop.
\appendix

\bibliography{DGDM}{}
\bibliographystyle{apsrev4-1}

\end{document}